\documentclass[sigconf]{acmart}
\renewcommand\footnotetextcopyrightpermission[1]{} 
\usepackage{multicol}
\usepackage{multirow}
\usepackage{algorithm}
\usepackage{algorithmic}
\usepackage{subfigure}
\usepackage{graphicx}
\begin{document}
\title{Unified Matrix Factorization with Dynamic Multi-view Clustering}

\author{Shangde Gao}
\authornote{Both authors contributed equally to this research.}
\affiliation{%
  \institution{Zhejiang University}
  \streetaddress{Zijinghua Road}
  \city{Hangzhou}
  \state{Zhejiang}
  \country{China}
  \postcode{310027}
}
\email{gaosde@zju.edu.cn}
\orcid{0009-0007-9831-3329}

\author{Ke Liu}
\authornotemark[1]
\affiliation{%
  \institution{Zhejiang University}
  \streetaddress{Zijinghua Road}
  \city{Hangzhou}
  \state{Zhejiang}
  \country{China}
  \postcode{310027}
}
\email{lk2017@zju.edu.cn}

\author{Yichao Fu}
\affiliation{%
  \institution{Zhejiang University}
  \streetaddress{Zijinghua Road}
  \city{Hangzhou}
  \state{Zhejiang}
  \country{China}
  \postcode{310027}
}
\email{fuyichao@zju.edu.cn}

\begin{abstract}

Matrix factorization (MF) is a classical collaborative filtering algorithm for recommender systems. 
It decomposes the user-item interaction matrix into a product of low-dimensional user representation matrix and item representation matrix. 
In typical recommendation scenarios, the user-item interaction paradigm is usually a two-stage process and requires static clustering analysis of the obtained user and item representations. 
The above process, however, is time and computationally intensive, making it difficult to apply in real-time to e-commerce or Internet of Things environments with billions of users and trillions of items. 
To address this, we propose a unified matrix factorization method based on dynamic multi-view clustering (MFDMC) that employs an end-to-end training paradigm. 
Specifically, in each view, a user/item representation is regarded as a weighted projection of all clusters. The representation of each cluster is learnable, enabling the dynamic discarding of bad clusters.
Furthermore, we employ multi-view clustering to represent multiple roles of users/items, effectively utilizing the representation space and improving the interpretability of the user/item representations for downstream tasks. 
Extensive experiments show that our proposed MFDMC achieves state-of-the-art performance on real-world recommendation datasets. 
Additionally, comprehensive visualization and ablation studies interpretably confirm that our method provides meaningful representations for downstream tasks of users/items.
\end{abstract}

\maketitle
\fancyfoot{}
\thispagestyle{empty}
\section{Introduction}
Recommender system is in great demand nowadays with the rapid growth of various web services, \emph{e.g.}, e-commerce and social network \cite{yu2023self,ko2022survey}. It helps users find the items of their interest from a massive amount of candidates, making significant contributions in improving user experience as well as increasing business value.

Matrix Factorization (MF) \cite{web:mf} is a classic and effective method for extracting user preferences (user latent vectors) and item features (item latent vectors) from historical data, which performs well in recommendation systems. 
In its basic form, the MF model, such as singular value decomposition, maps both users and items to a fixed dimensional joint latent factor space, where the user-item interactions are modeled as inner products in this space. 
The recommendation is formed by the highly corresponding item and user factors, represented by the inferred factor vectors from the item rating patterns. 
However, the traditional MF process requires significant time and computational resources, making it difficult to apply in real scenarios, especially when dealing with billions of users and tens of billions of items in big data situations.

Recently, many variants of MF have been proposed to deal with specific problems in recommendation systems.
For example, biased MF \cite{biased_mf} handles biases in ratings. 
Some deep learning (DL) methods ,otherwise, utilize deep neural networks to construct high-order features or model latent non-linear connections, achieving the generalization and improvement of matrix factorization algorithms for recommendation performance \cite{lian2018xdeepfm,zhu2022rating,liu10191051,gao2023contrastive}.

However, complex optimization designs were required for matrix factorization in previous works. Furthermore, the problem of the MF process being a black box, where the results are uninterpretable, has not been solved \cite{survey1}. Additionally, the latent representation space is often not fully and effectively utilized, resulting in high dimensionality and redundant information in the latent space.




In this work, we endeavor to explore an effective and interpretable MF scheme for recommder system. Our goal is to maximize the utilization of the latent representation space in acquiring user user/item representations, while significantly reducing time and computing resources. Additionally, we aim to explore the interpretability of user/item representations as much as possible and apply it to downstream specific tasks. Herefore, we propose an unified \textbf{M}atrix \textbf{F}actorization with \textbf{D}ynamic \textbf{M}ulti-view \textbf{C}lustering (MFDMC). The contributions of this work are summarized as follows: 
\begin{itemize}
    \item 
    We propose a unified matrix factorization method for recommender system, known as MFDMC. MFDMC combines dynamic clustering and matrix decomposition to fully leverage the representation space, resulting in significant time and computational resource savings.
    \item We conduct comprehensive visualization to validate the interpretability of user/item representations obtained from multi-view clustering.
    \item Extensive experiments on six real-world datasets demonstrate that MFDMC achieves state-of-the-art performance and can be applied to downstream computer vision tasks with constrained representation space.
\end{itemize}

\begin{table}[t]
    \centering
    \begin{tabular}{c|p{0.31\textwidth}}
         \hline
         Notations & Meaning \\
         \hline
         $d$ & Dimension of latent vector, $d \in \mathbb{Z}$\\
         $m, n$ & Number of users/items, $m, n \in \mathbb{Z}$\\
         $P, Q$ & Matrix of users/items, $P, Q \in \mathbb{R}^{m\times d}$\\
         $p_i, q_i$ & The $i^{th}$ user/item latent vector, $p_i, q_i\in \mathbb{R}^d$\\
         $R$ & User-item interaction matrix, $R \in \mathbb{R}^{m \times n} $\\
         $t,v$ & Number of centers, views, $e, v \in \mathbb{Z}$\\
         $b$ & Dimension of centers, $b \in \mathbb{Z}$ and $b = d/v$ \\
         $C^{user}, C^{item}$ & The centers of user/item, $C \in \mathbb{R}^{v \times e \times b}$ \\
         $c_{i,j}^{user}, c_{i,j}^{item}$ & The $i^{th}$ center in $j^{th}$ view of user/item, $c_{i,j} \in \mathbb{R}^{b}$\\
         $W^{user}_i, W^{item}_i$ & Weight of $i^{th}$ user/item for centers, $W \in \mathbb{R}^{v\times e}$\\
         $w_{i,j}^{user}, w_{i,j}^{item}$ & Weight of user/item $i^{th}$ center in $j^{th}$ view, $w_{i,j} \in \mathbb{R}$\\
         $\rho, \eta, \gamma, \psi, \lambda$ & Weight parameters\\
         \hline
    \end{tabular}
    \caption{The notations used throughout the paper}
    \label{tab:my_label}
    \vspace*{-0.4in}
\end{table}
\section{Related Works}
Matrix Factorization has garnered significant attention due to its efficacy in recommender systems, resulting in numerous studies focused on optimizing and improving its interpretability \cite{survey1, survey2, survey3, explainable}. Additionally, multi-view learning is extensively employed to enhance the performance of models. This section provides a brief overview of these techniques.

\subsection{Matrix Factorization}
The Matrix Factorization (MF) algorithm evolves from Singular Value Decomposition (SVD), which can only decompose dense matrices. However, the user-item interaction matrix is usually extremely sparse. Therefore, in order to decompose the rating matrix using SVD, the missing values of the matrix must be filled first. 
This can cause the following two problems: (1) Filling missing data can significantly increase the amount of data and lead to an increase in the complexity of the algorithm. 
(2) Improper padding methods can result in data distortion.

Since the SVD algorithm does not perform well with scoring matrices, researchers have turned to investigating whether they can decompose matrices considering only the existing scores. Matrix decomposition optimization methods such as FunkMF \cite{web:mf}, Probabilistic Matrix Factorization (PMF) \cite{pmf}, and BiasedMF \cite{biased_mf} have been proposed. Concretely, Simon Funk\cite{web:mf} introduced stochastic gradient descent to optimize Eq. \ref{equ:funkmf}. This algorithm traverses all rating records in the training set. For each training sample, the algorithm predicts the rating using the user/item embedding and computes the prediction error.
\begin{equation}
    \begin{aligned}
    \label{equ:funkmf}
    \min_{p,q} \quad \sum_{(u,i)\in \mathcal{K}}(r_{ui}-q_i^{T}p_u)^2 + \lambda(\Vert q_i \Vert ^2 + \Vert p_u \Vert ^2)
    \end{aligned}
\end{equation}

And BiasedMF \cite{biased_mf} takes individual biases into account, i.e., much of the observed variation in ratings is influenced by the user or item independently, rather than by the interaction between the user and the item. So BiasedMF divides the rating into four parts: the item and user biases, the interaction between them and global average as Eq. \ref{equ:biasedmf}. Where $b_i$, $b_u$, and $\mu$ are the user, item bias and the global average respectively.
\begin{equation}
    \begin{aligned}
    \label{equ:biasedmf}
    \hat{r_{ui}}= q_i^{T}p_u + b_i + b_u + \mu
    \end{aligned}
\end{equation}

Pu et al. proposed a model-based method called Asymmetric SVD \cite{pu} to tackle the cold-start problem. Bi et al. also proposed a method called Group-specific SVD \cite{Bi}, where users/items are grouped into clusters and their embeddings are influenced by the cluster they belong to. 

In recent years, the rapid development of deep learning has led to the integration of deep learning methods into Matrix Factorization (MF)\cite{dl1,dl2}. This integration aims to generate representational information by constructing higher-order features or potential nonlinear connections of models \cite{gao2023contrastive,liu10191051}, with the goal of improving both model performance and generalization ability.
\subsection{Multi-view Learning}
The idea of Multi-view learning \cite{wen2022survey, ren2022deep,liu2023s2snet} is widely utilized in various models such as multi-head attention in Transformer \cite{head1} and multi-interest in Multi-Interest Network with Dynamic routing (MIND) \cite{head2}. In the case of Transformer, multi-head attention allows the model to simultaneously focus on information from different representations of the information subspace in different locations. Similarly, in MIND, Li et al.  \cite{head2} employ multiple vectors to represent a user, thereby encoding different aspects of the user's interests. Both of these works demonstrate the effectiveness of the multi-view approach. In our method, we extend the concept of multi-view learning to an environment, where we design dynamic multi-view clustering to represent multiple roles of users/items, effectively leveraging the representation space and subsequently applying them to downstream recommendation tasks.
\subsection{Interpretability in Recommender Systems}
In the Matrix Factorization (MF) technique, the representation vectors of users and items are embedded in a low-dimensional space. Each dimension of this space represents a specific factor that influences the user's decision. However, the precise meaning of each factor is unclear, making it difficult to interpret predictions or recommendations. 

Several interpretable recommendation models have been proposed based on matrix factorization methods \cite{explainable}. 
In the Explicit Factor Models (EFM) \cite{emf} approach, Zhang et al. extract explicit features from user reviews and assign each latent dimension to a specific feature. 
This makes the entire process traceable and provides an explicit interpretation. 
In another approach, Chen et al. construct a user-item-feature cube by extracting features from user reviews \cite{explanation1}. 
They then employ a pair-wise learning-to-rank method to rank the features and items. 
The Sentiment Utility Logistic Model (SULM) \cite{expl2}, presented by Bauman et al., incorporates user sentiments on these features into MF to predict ratings. 
This enables the learning of feature recommendations for each item, which can be used as explanations. In the Explainable Matrix Factorization (EMF) 
 method \cite{expl3}, an "interpretable regularizer" is added to the objective function of MF. 
This generates relevant-user interpretations. 

Among the aforementioned methods, only EFM offers dimension-wise interpretability. 
Additionally, all of them require extra information beyond the user-item interaction matrix, except for EMF. However, both EFM and EMF underperform in terms of Root Mean Square Error (RMSE), which is an important metric in experiments.
\begin{figure}[t!] \includegraphics[width=\columnwidth]{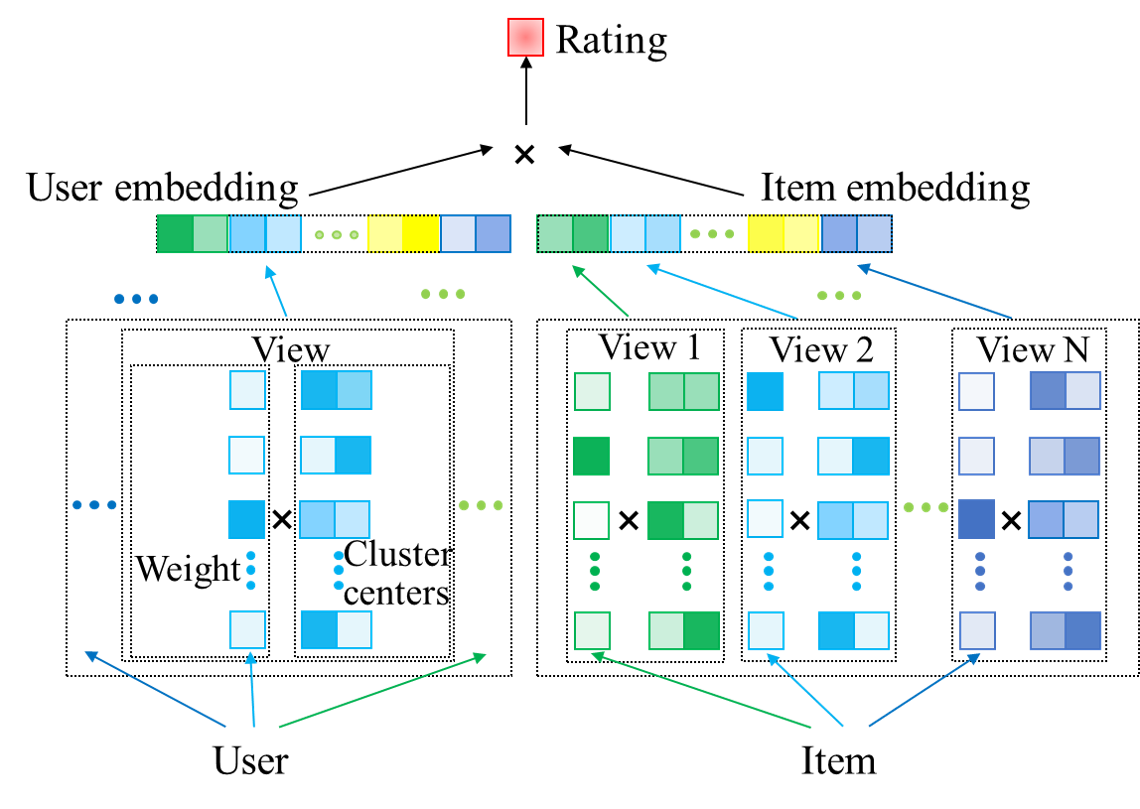}
    \caption{Illustration of our MFDMC. The User/item is embedded into 
    Weights $t$ centers in $v$ Views. In each view, the representation is weighted sum of the Cluster centers.
    }\label{fig:model}
    \vspace*{-0.35in}
\end{figure}
\section{MF with Dynamic Multi-view Clustering}
The purpose of matrix decomposition is to characterize users and items by decomposing the user-item interaction matrix into user and item matrices in a lower-dimensional latent space, as formulated in Eq. \ref{equ:orig}. 
Each row of matrix $P$ or $Q$ represents a latent vector for user or item representation.
\begin{equation}
\label{equ:orig}
R = P \cdot Q^T
\end{equation}

In this section, we propose a unified matrix factorization with Dynamic Multi-view Clustering (MFDMC), in a end-to-end training paradigm, to obtain interpretable representations and make full use of the representation space. The overview of MFDMC is shown in Fig. \ref{fig:model}, where the user/item representation provides rich information about users' interest preferences, such as preferences for comedy or the perceived humor in movies. Therefore, latent vectors are divided into sub-vectors in $v$ views of interest. In each view of interest, the user/item is assigned to one of the clusters, which represents the preference, and the cluster center is selected. This way, the user/item latent vector can be derived as shown in Eq. \ref{equ:sub1}.\begin{equation}\begin{aligned}\label{equ:sub1}p_i = \bigoplus_{j=0}^{v} \sum_{i=0}^t c_{i,j}^{user} w_{i,j}^{user} \\q_i = \bigoplus_{j=0}^{v} \sum_{i=0}^t c_{i,j}^{item} w_{i,j}^{item}\end{aligned}\end{equation}The numbers of views $v$ for users and items are the same, while the center numbers $e$ can vary. 

Loss functions are designed for cluster centers, user/item weights, and target user-item interaction ratings to achieve the aforementioned goals, which are detailed in the subsections.

\subsection{Cluster centers}
In this section, our clustering method aims to address the following two problems: (1) How to fully utilize the latent space in the cluster centers? (2) How to manage the number of clusters during training? Firstly, the latent space is expected to be fully utilized, which means that the cluster centers should be spread out. We design the loss function as shown in Equation \ref{equ:loss1u}:\begin{equation}\begin{aligned}\label{equ:loss1u}loss_1^{user} =  \sum_{j=0}^{v} \sum_{\alpha,\beta}\mathit{l}(c_{\alpha,j}^{user}, c_{\beta,j}^{user})\\loss_1^{item} = \sum_{j=0}^{v} \sum_{\alpha,\beta}\mathit{l}(c_{\alpha,j}^{item}, c_{\beta,j}^{item})\\\mathit{l}(c_{\alpha}, c_{\beta}) = max\{0, \rho - \mathcal{D}(c_{\alpha} - c_{\beta})\}\end{aligned}\end{equation}Here, $\mathcal{D}$ represents any distance function, and $\rho$ defines the maximum allowed proximity between cluster centers. Additionally, to bring the latent vectors closer to their corresponding cluster centers, we measure the average distance between a user/item and its cluster center using Equation \ref{equ:meanu}:

\begin{equation}\label{equ:meanu}
\begin{aligned}
    loss_{1,c}^{user}=\frac{1}{N}\sum_{j=0}^{v} \sum_{i=0}^{t}\sum_{k=0}^{N,k \in S_{i,j}}\Vert c_{i,j}^{user}-p_k^{user}\Vert^2 \\
    loss_{1,c}^{item}=\frac{1}{N}\sum_{j=0}^{v} \sum_{i=0}^{t}\sum_{k=0}^{N,k \in S_{i,j}}\Vert c_{i,j}^{item}-q_k^{item}\Vert^2 \\
\end{aligned}
\end{equation}
where $S_{i,j}$ is the set of user/item that belong to the $i^{th}$ cluster in the $j^{th}$ view.

Then, to make full use of the representation space, the loss function we want to optimize for cluster centers can be defined as:
\begin{equation}
    \label{equ:loss1a}
    loss1 = loss_1^{user}+loss_1^{item}+loss_{1,c}^{user}+loss_{1,c}^{item}
\end{equation}

Considering the difficulty of determining the number of cluster centers that accurately represent potential space, it is also challenging to ascertain if there is redundancy in our problem. In our MFDMC, we employ a dynamic update mechanism to update and manage the cluster centers' count. The Algorithm \ref{alg:alg1} provides an outline on how to handle cluster centers while calculating cluster losses. Specifically, after each iteration of $I_p$, we calculate the weights of each center and remove those whose weights fall below the threshold $\psi$.
\begin{algorithm}[t!]
\caption{Dynamic clustering algorithm}
\label{alg:alg1}
\textbf{Input}: Cluster centers: $\mathcal{C}$; Weights of user/item: $\mathcal{W}$; Current epoch: $i$.\\
\textbf{Output}: New cluster centers: $\mathcal{C}'$; Clustering loss: $\mathcal{L}_{1,c}$
\begin{algorithmic}[1] 
\IF {$i > I_p$}
\STATE $\overline{\mathcal{W}} \leftarrow $ cluster-wise mean of user/item in $W$.
\STATE $\mathcal{C}',\mathcal{W}\leftarrow $ Remove the $\mathcal{C}$ and $\mathcal{W}$ where $\overline{\mathcal{W}} < \psi$.
\ELSE
\STATE $\mathcal{C}' \leftarrow \mathcal{C}$.
\ENDIF
\STATE $\mathcal{P},\mathcal{Q} \leftarrow \mathcal{W}$ weighted sum $\mathcal{C}'$.
\STATE $\mathcal{L}_{1,c} \leftarrow$ Compute using $\mathcal{C}'$, $\mathcal{P}$, $\mathcal{Q}$ and Eq.\ref{equ:meanu}.
\STATE \textbf{return} $\mathcal{C}'$, $\mathcal{L}_{1,c}$
\end{algorithmic}
\end{algorithm}
\begin{figure}[b!]
\vspace*{-0.2in}
   \includegraphics[width=\columnwidth]{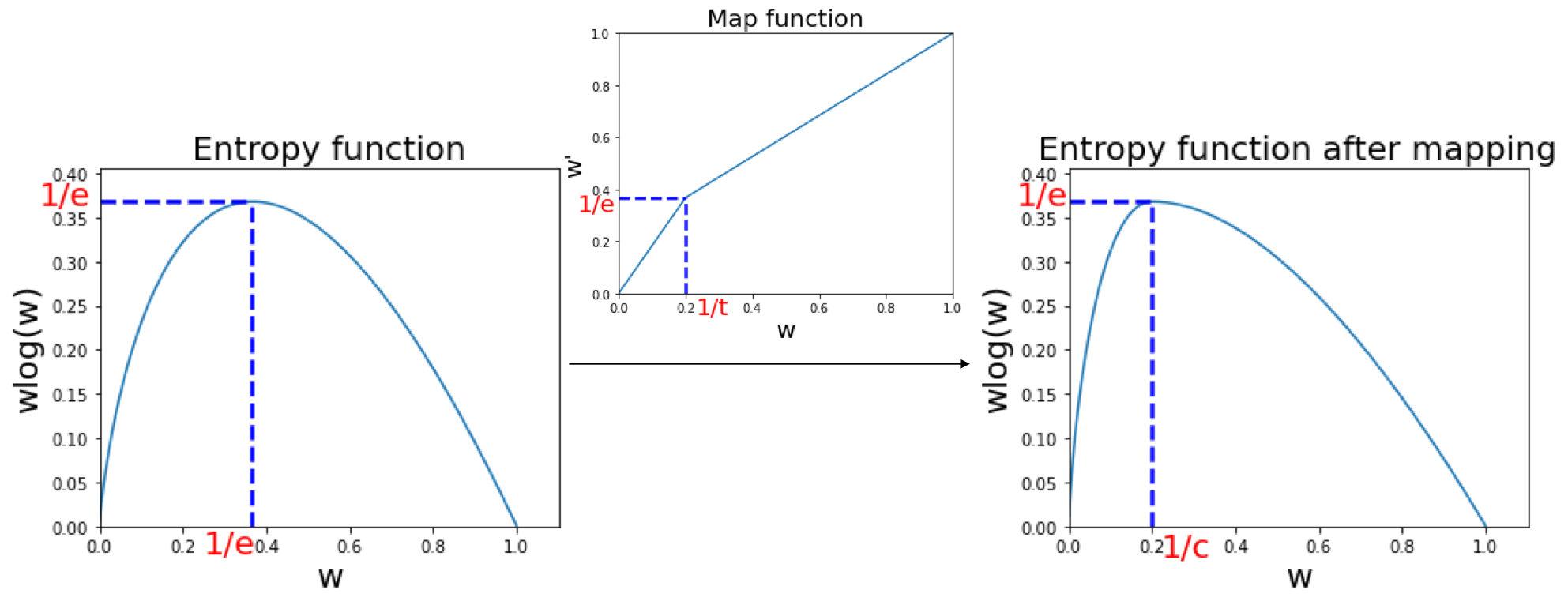}
    \caption{Mapping function of Eq. \ref{equ:map} for the weight. Using the function in the middle of the Figure, we can map the weights on the left to the right, which facilitates a relatively balanced optimization of weight in different views.
    }\label{fig:map}
\end{figure}
\subsection{User/item weights}
From the perspectives of optimization and interpretability, user/item should be explicitly grouped into clusters in an interest view. Therefore, the $softmax$ function is used to normalize the weights in Equation \ref{equ:loss21}, and entropy is used in the loss function \ref{equ:loss23} to reduce uncertainty.
\begin{equation}
\begin{aligned}
    \label{equ:loss21}
    W^{user'} = softmax(W^{user}) \\
    W^{item'} = softmax(W^{item}) 
\end{aligned}
\end{equation}

\begin{table}[t!]
\centering
\small
\begin{tabular}{p{2.2cm}|p{0.9cm}p{0.85cm}p{1.05cm}p{1.3cm}}
    \hline
    Dataset & $m$ & $n$ & $N$ & Range\\
    \hline
    MovieLens-1M & 6,040 & 3,952 & 1,000,209 & [1, 5] \\ 
    MovieLens-100k & 943 & 1,682 & 100,000 & [1, 5]  \\
    Amazon-video & 424,560 & 23,745 & 583,933 & [1, 5]  \\ 
    Epinions & 40,163 & 139,738 & 664,824 & [1, 5]  \\ 
    Books-across & 105,283 & 340,395 & 1,149,780 &[0, 10] \\ 
    Jester & 73,421 & 100 &	3,519,446 & [-10, 10]\\ 
    \hline
\end{tabular}
\caption{Statistics of the datasets. The symbol $N$ indicates the number of records. And $Range$ indicates the range of ratings.}
\label{dataset}
\vspace{-0.4in}
\end{table}

Additionally, the number of centers varies from view to view. In views with a different number of centers, the losses behave differently even if the weight distribution is the same. For example, in $view_1$ and $view_2$, there are 3 and 10 centers respectively, and the weight distribution of users is uniform. Although the distribution of both views is poor, the loss of $view_1$ is almost at the extreme point and much greater than the loss of $view_2$. This can lead to unbalanced optimizations, meaning that it is much more difficult to optimize losses in views with more centers. Therefore, the mapping function of $w$ in Equation \ref{equ:loss23} is designed to solve this problem. As shown in Fig. \ref{fig:map}, the uniform distribution loss (worst case) is always the extreme value even in views with a different number of centers, where $e$ is the Napier logarithm.


\begin{equation}
    \label{equ:map}
    w''=\left\{\begin{array}{ccl}
        \frac{t}{e}w' && 0 \leq w' \leq \frac{1}{t} \\
        \frac{tw'-1}{t-1}(1-\frac{1}{e})+\frac{1}{e} && \frac{1}{t}\le w' \le 1
    \end{array} \right.
\end{equation}
\begin{equation}
\label{equ:loss23}
\begin{aligned}
    loss_2^{user} = -\sum_{j=0}^{v} \sum_{i=0}^{m} w_{i,j}^{user''}log(w_{i,j}^{user''}) \\
    loss_2^{item} = -\sum_{j=0}^{v} \sum_{i=0}^{n} w_{i,j}^{item''}log(w_{i,j}^{item''}) 
\end{aligned}
\end{equation}

Finally, the loss for user/item weight is taken to be: 
\begin{equation}
\label{equ:loss2}
    loss_2 = loss_2^{item} + loss_2^{user}
\end{equation}

\textbf{Evaluation metric} Root mean square error (RMSE) \ref{equ:rmse} is used as evaluation metric and one of the loss.
\begin{equation}
    \label{equ:rmse}
    loss_3 = \sqrt{\frac{1}{N}\sum_{i=0}^{N}(y_i - r_i)^2}
\end{equation}

Finally, the total objective of MFDMC can be described as:
\begin{equation}
    \label{equ:loss}
    loss = \eta loss_1 + \gamma loss_2 + loss_3
\end{equation}
\begin{table*}[h]
\centering
\begin{tabular}{llllllll}
    \hline
    Method & MovieLens-1M & MovieLens-100k & Amazon-video & Epinions & Books-across & Jester\\
    \hline
    FunkMF(16) &  0.869  & 0.938 & 2.063 & 1.637& 3.836 & 4.973\\
    FunkMF(60) & 0.868  & 0.936 & 2.163 & 1.776 & 3.863 & 5.049\\
    \hline
    PMF(60) & 0.883 & 0.952&- &- &-&-\\
    BiasedMF(16) & 0.866 & 0.926 & 1.034 & 1.039 & 3.508 & 4.904\\
    BiasedMF(60) & 0.863 & 0.923 & 1.030 & 1.044 & 3.503 & 4.906\\
    \hline
    Ablation(16) & 0.862  & 0.943 &1.036 & 1.043 & 3.398 & 4.220\\
    MFDMC(16) & \textbf{0.848} & \textbf{0.911}& \textbf{1.019} & \textbf{1.034} &  \textbf{3.352} & \textbf{4.114}\\
    \hline
\end{tabular}
\caption{RMSE across the two data sets for a variety of techniques. The token (D) specifies the dimension of the latent vector $d$. RMSE reported for PMF is taken from \cite{result}. Scores for BiasedMF and FunkMF are obtained using surpriselib \cite{surprise}.}
\label{table:result}
\vspace*{-0.2in}
\end{table*}

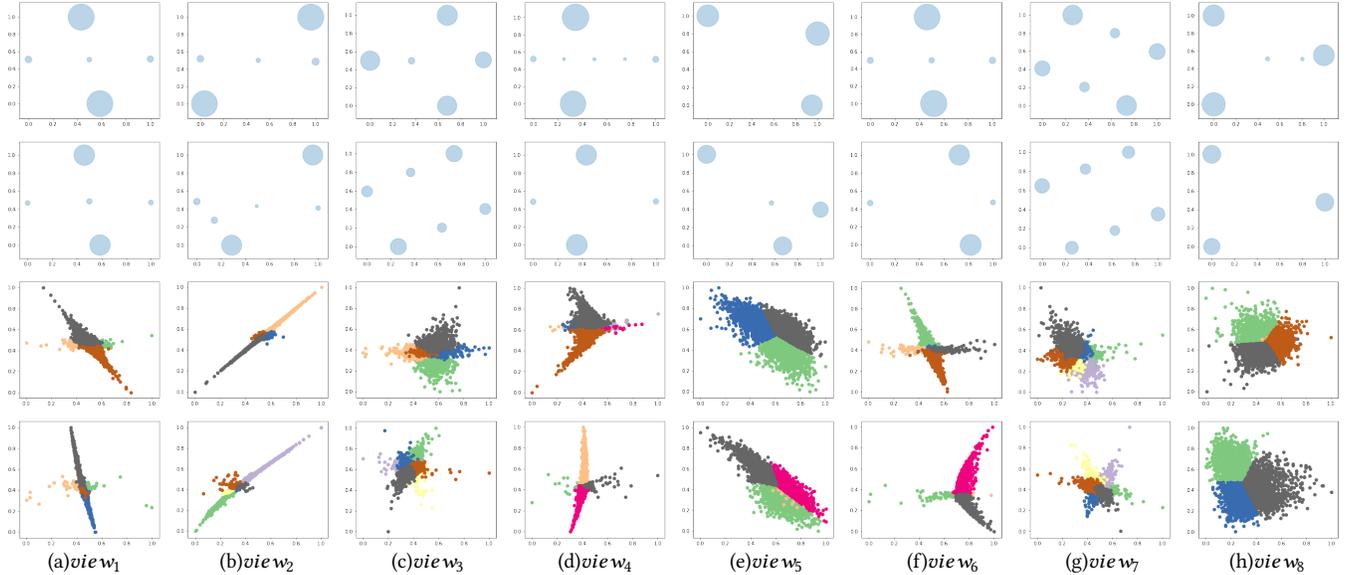
\begin{figure*}[t!]
  \includegraphics[width=\linewidth]{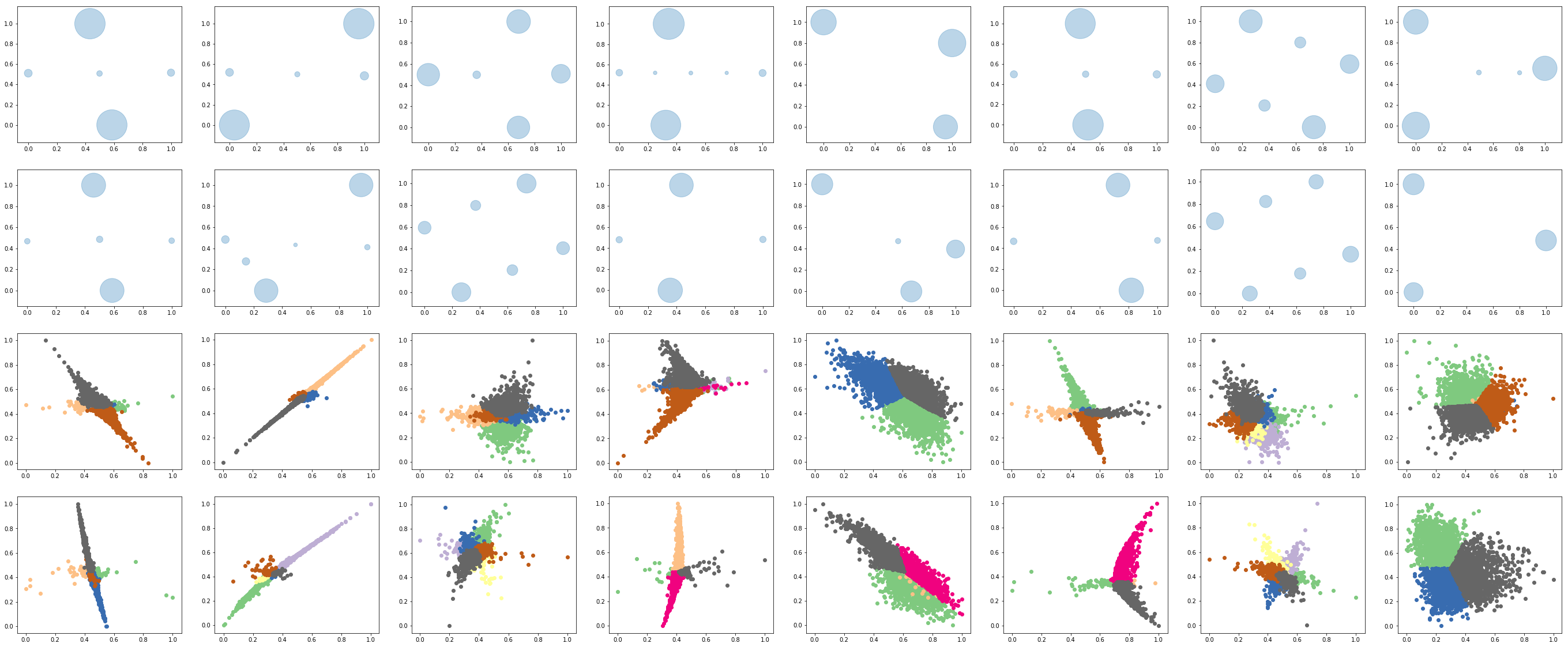}\small
    \put(-490,-7){(a)$view_1$}
    \put(-425,-7){(b)$view_2$}
    \put(-360,-7){(c)$view_3$}
    \put(-297,-7){(d)$view_4$}
    \put(-232,-7){(e)$view_5$}
    \put(-165,-7){(f)$view_6$}
    \put(-105,-7){(g)$view_7$}
    \put(-43,-7){(h)$view_8$}
  \caption{Multi-view Clustering results. The four rows show the clustering centers of users/items, the clustering results of users/items respectively. Each column represents a view. In the clustering result, each point is the value of user/item latent vector in one view. And the size of cluster centers indicates the number of user/item in the cluster }\label{fig:cluster}
  \vspace*{-0.2in}
\end{figure*}
\section{Experiments}
In this section, we present extensive experimental results on six real-world datasets. 
We conduct a thorough comparison with matrix factorization and visualization techniques to validate the superiority and interpretability of our MFDMC model.
\subsection{Datasets \& Implementation Details}
\subsubsection{Datasets}
The experiments are conducted on six datasets, each of which contains files of users, movies, and ratings. Only the rating files are used in the experiments. An overview of the datasets is provided in Table \ref{dataset}.For all the datasets, we randomly select 80\% of the data for training the model, 10\% of the data for validation, and report the results of the remaining 10\% of the data for testing.
\textit{MovieLen} \cite{dataset:movielens} collects user ratings for movies. Meanwhile, for each user in the MovieLens 1M and 100k, there are at least 20 ratings. \textit{Amazon-video} \cite{dataset:amazon} is a subset of Amazon review dataset, in which each rating indicates how the user rated the instant video. Each user rated at least one video. \textit{Epinions}\footnote{\url{http://www.trustlet.org/datasets/downloaded_epinions/}} collects subjective reviews of users about many different types of items. Also, at least one rating exists for a user. \textit{Books-across} \cite{dataset:books} contains user's more than one rating on the books. \textit{Jester} \cite{dataset:jester} collects the ratings on jokes by users, where each user has rated more than 15 jokes.
\subsubsection{Parameter Setting of MFDMC}
As stated in Eq. \ref{equ:loss1u}, $\mathcal{D}$ is used to measure the distance between user/item and its cluster center. In our experimental setup, we use the Euclidean distance as the metric, as shown in Eq.\ref{equ:loss1d} below
\begin{equation}
    \label{equ:loss1d}
    \mathcal{D}(c_{\alpha}, c_{\beta}) = \Vert c_{\alpha} - c_{\beta} \Vert^2
\end{equation}
Furthermore, cluster centers are view-wise normalized by Eq.\ref{equ:norm}
\begin{equation}
    \label{equ:norm}
    C = \frac{C-min(C)}{max(C)-min(C)}
\end{equation}
In our all experiments, $I_d$ is set to be 40, which means, in the first 40 epochs, the cluster centers are not removed dynamically. The threshold to remove the cluster center is $\frac{1}{t}$ and we keep at least 3 centers in one view. The weight parameters, $\eta$ and $\gamma$ of $loss_1$ and $loss_2$ increase gradually with the epochs. As the numbers of views for users and items are the same, centers in the same location in the view can be shared. And the number of cluster centers $t$ in each view is 10. Weight decay with a regularization parameter $\lambda$ is also added to our model.
\subsection{Results \& Analysis}
\subsubsection{Results}
Firstly, we compared our approach with other baseline models. Table~\ref{table:result} presents the RMSE values for four methods on extensive datasets. 
Three observations were made. 
(1)MFDMC consistently outperformed the other competitors, such as achieving approximately 0.025 RMSE improvement compared to FunkMF on the MovieLens-100k dataset. 
(2) Despite the increase in the dimension of the latent space from 16 to 60 in FunkMF and BiasedMF, the RMSE did not improve or even worsened. These models were unable to fully utilize the representation space.
However, in MFDMC, where the dimension of the latent space is 16 (same as FunkMF), the RMSE was significantly better. 
It was even observed that MFDMC with a lower dimension of 12 (as shown in Table~\ref{table:config}) could achieve comparable results to other methods.
(3) The ablation studies revealed that our approach was effective not only due to the deep structure of our model but also because of the loss functions we designed. 
Without these loss functions, the RMSE of our model was only slightly better than or even worse than the other methods.
However, with the inclusion of the loss functions, the RMSE consistently improved.

Besides, our approach dynamically clusters users and items from various perspectives by decomposing the interaction matrix. To provide a clear visualization of the clustering results, we utilized the results of experiment No.4 in Table~\ref{table:config}. As depicted in the Fig.~\ref{fig:cluster}, every user and item is assigned to a specific cluster. 
Although the cluster boundaries may not be distinct, we can readily determine the cluster to which each user or item belongs based on their weights assigned to each center.
\subsubsection{Ablation analysis}
\begin{figure}[t]
\includegraphics[width=0.7\columnwidth]{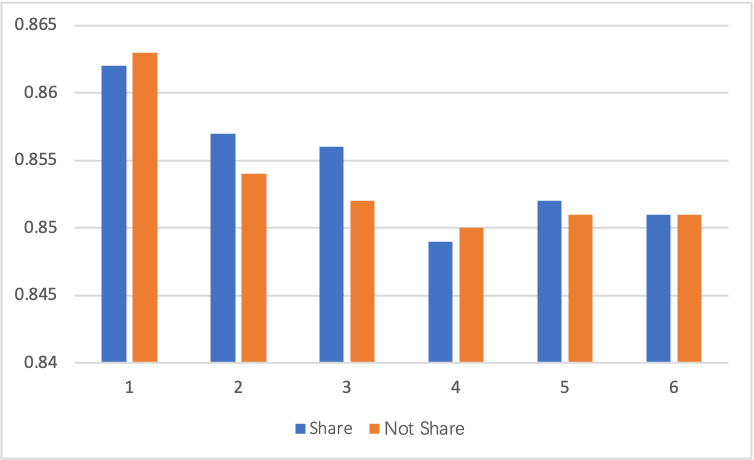}
    \caption{ Performance (RMSE) comparison of applying different configurations. The horizontal and vertical coordinates are the experiment number and RMSE  respectively.
    }\label{fig:config}
    \vspace*{-0.25in}
\end{figure}

\begin{table}[h]
\centering
\begin{tabular}{llllll}
    \hline
    No. & Not Share & Share & $v$  & $d$ & $b$\\
    \hline
    1 & 0.862 & 0.863 &2 & 12 & 6 \\
    2 & 0.857 & 0.854 & 4 & 12 & 3 \\
    3 & 0.856 & 0.852 & 6 & 12 & 2 \\
    4 & 0.849 & 0.850 & 8  & 16 & 2 \\
    5 & 0.851 & 0.852 & 8 & 24 & 3 \\
    6 & 0.851 & 0.851 & 10 & 20 & 2 \\
    \hline
\end{tabular}
\caption{Performance (RMSE) comparison of applying different configurations.}
\label{table:config}
\vspace*{-0.3in}
\end{table}
\begin{figure*}[t]
\centering  
\subfigure[Color view]{
\label{Fig.sub.1}
\includegraphics[width=0.24\linewidth]{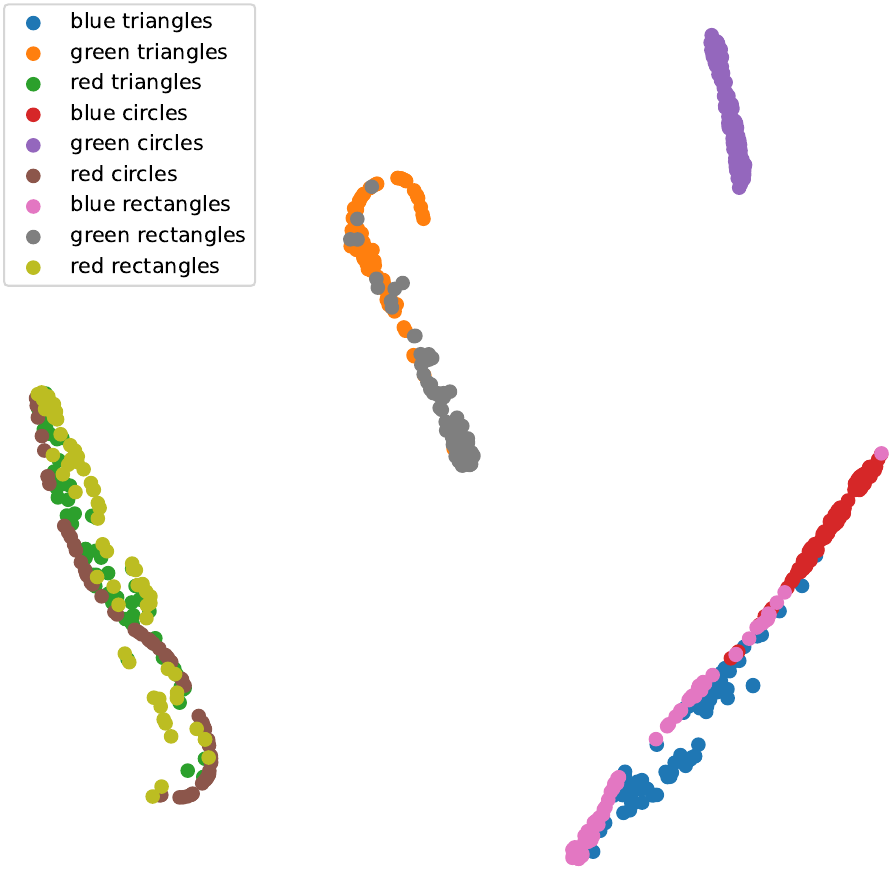}}
\subfigure[Shape view]{
\label{Fig.sub.2}
\includegraphics[width=0.24\linewidth]{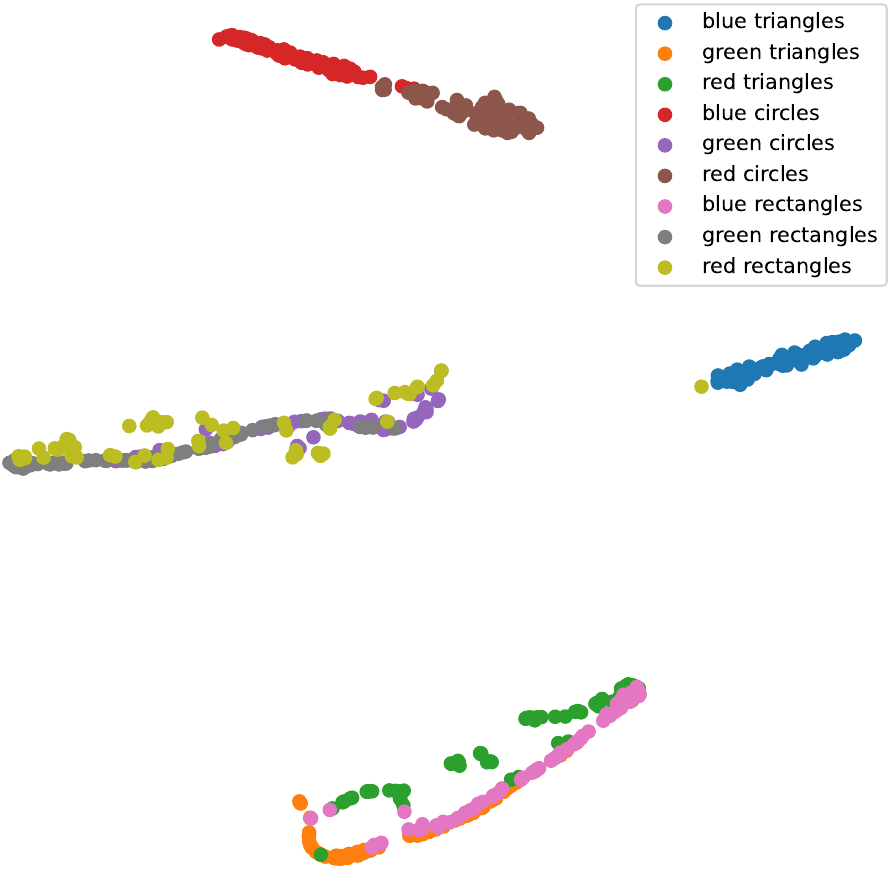}}
\subfigure[Entire embedding]{
\label{Fig.sub.3}
\includegraphics[width=0.24\linewidth]{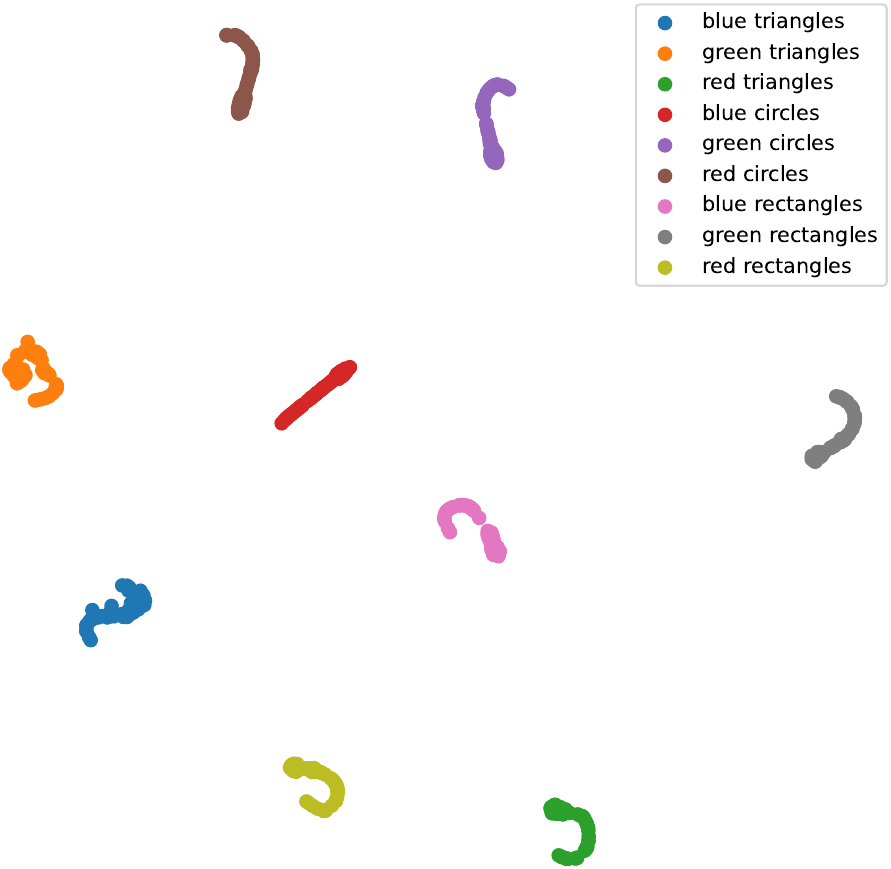}}
\caption{The t-sne results of experiments on CV. Fig.~\ref{Fig.sub.1}, Fig.~\ref{Fig.sub.2}, and Fig.~\ref{Fig.sub.3} show the t-sne results of shape view, color view, and the entire embedding respectively.}
\label{Fig.main}
\vspace*{-0.2in}
\end{figure*}
We conducted ablations using different set-ups on MovieLens-1M in order to determine the basic method of assigning the number of views, centers, etc. The results of these experiments, including the RMSE, can be found in Table~\ref{table:config} and Fig.~\ref{fig:config}.Based on the results, we observed that when the latent space dimension ($d$) is kept constant and the number of views ($v$) is increased, the average RMSE improves regardless of whether the center is shared or not. However, if the number of views becomes too large, the RMSE starts to worsen. Additionally, whether the center is shared or not does not have a significant impact on the experimental results. Similarly, increasing the dimension of the latent space ($d$) does not significantly improve the RMSE for the same number of views ($v$). Based on these observations, we suggest the following parameter settings for $v$, $d$, and $b$:1. Determine the appropriate number of views ($v$) based on the number of classes you want to cluster the items. In the case of MovieLens-1M, there are 16 categories of movies, and our goal was to group them into six main categories.2. If you need to save time and memory resources, you can share centers between users and items.3. Keep the dimension of centers ($b$) low, as a small $b$ is sufficient to adequately represent the properties of a cluster. For example, in the No.4 experiment, we found that $b=2$ worked well.

\subsubsection{Interpretability analysis}
\begin{figure}[t]  \includegraphics[width=0.7\columnwidth]{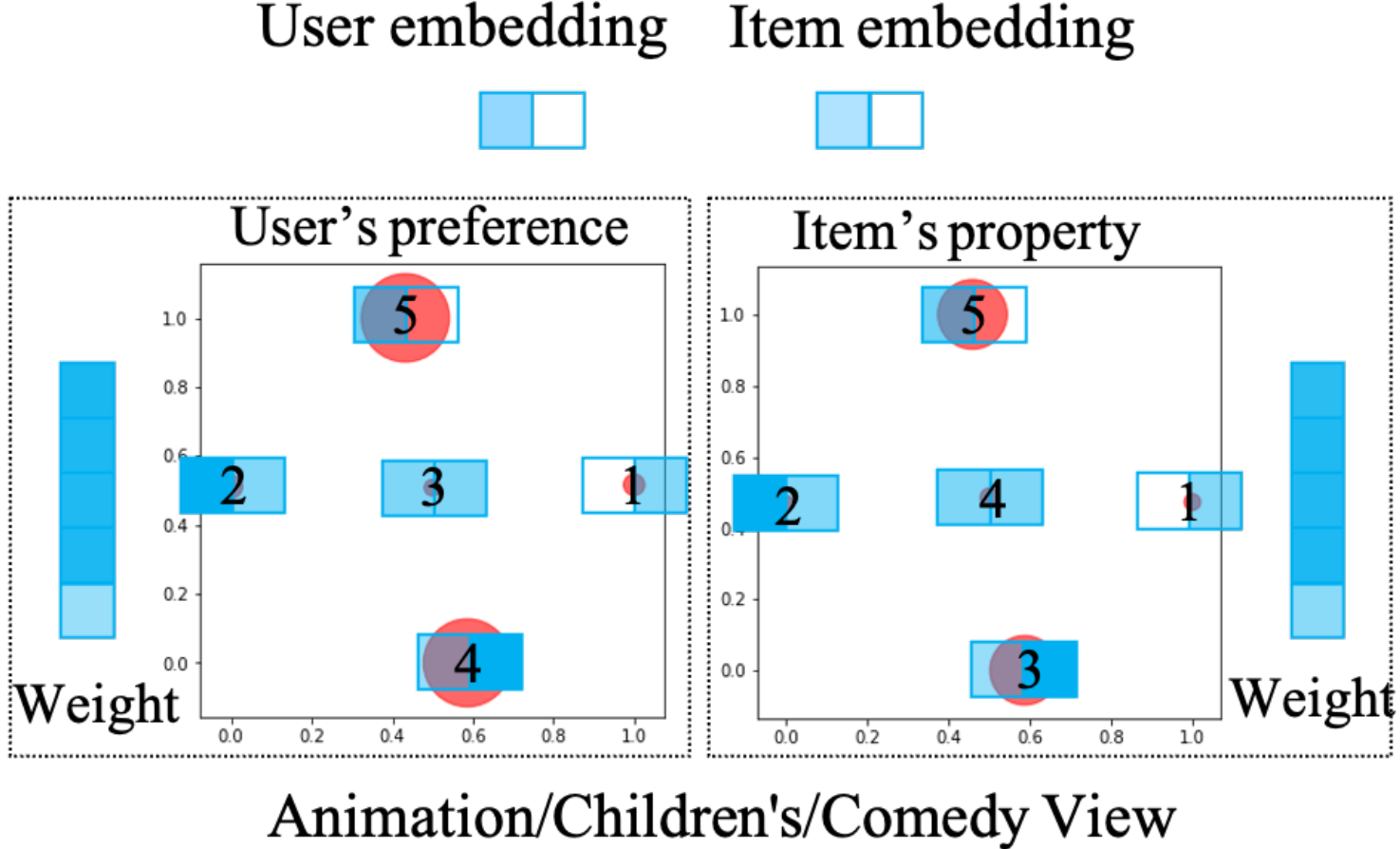}
    \caption{Illustration of interpreting a representation.  The red circles represent the cluster centers, while the squares depict the embedding of centers and weights. The transparency of each shape indicates its value, and the number on the center indicates the position of its corresponding weights.
    }\label{fig:illustration}
    \vspace*{-0.28in}
\end{figure}

We take one $view$ of the No.4 experiment in Table~\ref{table:config} as an example of how to interpret the user/item representation. 
Via counting the movies in the clusters of this view, it is clear that movies in the category of animation, children and comedy categories are located in several clusters, while movies in other categories are almost all in one cluster. Hence $view_1$ represents the movies in these categories. Specifically for any of the clusters in the view of user, by counting the user's rating on the movies in these categories, it can be inferred which cluster represents more preference and which cluster represents movies with higher quality in that category.
We randomly select a user and a movie to get their embedding respectively, which is obtained from the weighted sum of centers in $view_1$. The overview is shown in Fig.~\ref{fig:illustration}. For example, through the analysis above, the meaning of each center is obtained. Therefore, when a user/item belongs to a certain cluster, he also has the properties of that cluster, i.e., the category and quality of the movie or the user's preferences.

Moreover, additional experiments on CV were conducted to verify the interpretability. The results are shown in Fig.~\ref{Fig.main}. In the color view of Fig.~\ref{Fig.sub.1}, the green, red, and blue geometric shapes are grouped together respectively. In the shape view of Fig.~\ref{Fig.sub.2}, although some of the shapes are incorrectly clustered, the green rectangles and red rectangles, blue circles and red circles, green triangles and red triangles are correctly clustered together. Fig.~\ref{Fig.sub.3} shows the t-SNE results of the entire embedding, where the 9 classes of shapes are correctly clustered.
\section{Conclusion}
In this study, we propose a novel approach called Unified Matrix Factorization with Dynamic Multi-View Clustering (MFDMC). 
Our motivation stems from the observation that the representation space is not fully utilized in traditional matrix factorization (MF) algorithms, leading to uninterpretable user/item representations. Additionally, downstream clustering tasks require significant additional time and resources. To address these limitations, we introduce dynamic multi-view clustering to MF. 
By incorporating dynamic multi-view clustering into MF, our proposed method not only enhances the interpretability of representations but also optimally utilizes the representation space. 
We validate our approach through extensive experiments on massive datasets, and the results demonstrate that our proposed MFDMC surpasses existing MF methods. 
Moreover, the dynamic multi-view clustering approach that we introduce effectively utilizes the representation space and enhances the interpretability of representations. 

As for future work, we plan to further improve our approach and extend its applicability to a broader range of models.

\bibliographystyle{ACM-Reference-Format}
\bibliography{acmmm}
\end{document}